\documentclass[useAMS,usenatbib,usegraphicx]{mn2e}
\usepackage{times}
\usepackage{amsmath}
\usepackage{epstopdf}

\newif\ifAMStwofonts
\AMStwofontstrue

\newcommand{\simlt}{\lower.5ex\hbox{$\; \buildrel < \over \sim \;$}}
\newcommand{\simgt}{\lower.5ex\hbox{$\; \buildrel > \over \sim \;$}}
\newcommand{\be}{\begin{equation}}
\newcommand{\ba}{\begin{eqnarray}}
\newcommand{\ee}{\end{equation}}
\newcommand{\ea}{\end{eqnarray}}

\title[Age-dating the TFR at z$\sim$0.5]
{Age-dating the Tully-Fisher relation at moderate redshift
\thanks{Based on observations collected at the European Southern
  Observatory, Cerro Paranal, Chile (ESO Nos. 65.O-0049, 66.A-0547,
  68.A-0013, 69.B-0278B and 70.B-0251A) and observations with the
  NASA/ESA \emph{Hubble Space Telescope}, PID 9502 and 9908.}
}
\author[Ferreras et al.]
{Ignacio Ferreras$^1$\thanks{E-mail: i.ferreras@ucl.ac.uk},
Asmus B\"ohm$^2$, Bodo Ziegler$^3$, Joseph Silk$^4$\\
$^1$ Mullard Space Science Laboratory, University College London, 
Holmbury St Mary, Dorking, Surrey RH5 6NT\\
$^2$ Institute of Astro- and Particle Physics, Technikerstrasse 25/8,
6020 Innsbruck, Austria\\
$^3$ University of Vienna, Department of Astrophysics, 
T\"urkenschanzstr. 17, 1180, Wien, Austria\\ 
$^4$ Institut d'Astrophysique, Universit\'e Pierre et Marie Curie, 98 bis
Boulevard Arago, 75014 Paris, France}

\begin{document}
\date{\sl MNRAS: Accepted 2013 October 18. Received 2013 October 11; in original form
2013 April 05}
\pagerange{\pageref{firstpage}--\pageref{lastpage}} \pubyear{2013}
\maketitle
\label{firstpage}

\begin{abstract}
We analyse the Tully-Fisher relation at moderate redshift from the
point of view of the underlying stellar populations, by comparing
optical and NIR photometry with a phenomenological model that combines
population synthesis with a simple prescription for chemical
enrichment. The sample comprises $108$ late-type galaxies extracted
from the FORS Deep Field (FDF) and William Herschel Deep Field (WHDF)
surveys at z$\simlt$1 (median redshift z=0.45). A correlation is found
between stellar mass and the parameters that describe the star
formation history, with massive galaxies forming their populations
early ($z_{\rm FOR}\!\sim\!3$), with star formation timescales,
$\tau_1\sim\!4$\,Gyr; although with very efficient chemical enrichment
timescales ($\tau_2\!\sim\!1$\,Gyr). In contrast, the
stellar-to-dynamical mass ratio -- which, in principle, would track
the efficiency of feedback in the baryonic processes driving galaxy
formation -- does not appear to correlate with the model
parameters. On the Tully-Fisher plane, no significant age segregation
is found at fixed circular speed, whereas at fixed
stellar-to-dynamical mass fraction, age splits the sample, with older
galaxies having faster circular speeds at fixed M$_s$/M$_{\rm
  dyn}$. Although our model does not introduce any prior constraint on
dust reddening, we obtain a strong correlation between colour excess
and stellar mass.
\end{abstract}

\begin{keywords}
galaxies: evolution -- galaxies: formation -- galaxies: stellar content -- 
galaxies: fundamental parameters.
\end{keywords}

%%%%%%%%%%%%%%%%%%%%%%%%%%%%%%%%%%%%%%%%%%%%%%%%
\section{Introduction}

Scaling relations among independent observational properties reveal
the presence of important mechanisms underlying the formation and
evolution of galaxies, and in some instances the interplay
between the baryons and the dark matter halos where galaxies
reside. The strong correlation between luminosity and circular speed
in disc galaxies \citep[i.e. the Tully-Fisher relation,][hereafter
  TFR]{TF:77} has been used not only as an important rung in the
cosmological distance ladder \citep[e.g.][]{Giov:97}, but also as a
strong constraint on models of galaxy formation
\citep[e.g.][]{Mathis:02,Dutton:07}.  A variation of the TFR,
  defined between maximum circular velocity and stellar plus gas mass
  \citep[i.e., the baryonic TFR,][]{McGaugh:00} features a much lower
  scatter than the traditional TFR, and has even been used to test
the validity of standard Newtonian mechanics \citep{McGaugh:12}. The
redshift evolution of the TFR allows us to constrain the star
formation and assembly histories of galaxies \citep[see,
  e.g.,][]{PSL:07}. From the observational side, there are technical
challenges that make the interpretation of the velocity field and the
determination of M/L ratios rather complicated. After the first
results regarding the redshift evolution of the TFR
\citep[e.g.][]{Vogt:96}, a number of papers followed, claiming an
evolution of the TFR slope \citep{Ziegler:02,Bohm:04}, with low-mass
discs being more luminous in the the past. However, the analysis of
\citet{Weiner:06} showed, instead, a significant brightnening of
massive galaxies at high redshift, a result that could be reconciled
if the star formation histories of massive galaxies have shorter
timescales, as expected from multi-colour studies of disc galaxies
\citep[see, e.g.,][]{FSBZ:04}. Constraining the TFR in a robust way is
clearly a difficult task; the scatter of the relation is rather large
for the amount of evolution observed, the outcome depends sensitively
on the passband used, and samples at z$\sim$1 are inherently biased
towards the brightest galaxies \citep{FerLor:09}. Furthermore,
\citet{Bohm:07} demonstrated that an evolution of the TFR scatter can
mimic an evolution in slope. In addition, the complexity of disc
kinematics is difficult to disentangle with single slit measurements,
whereas 2D velocity fields from Integral Field Unit data suggest no
evolution of slope, intercept or scatter of the TFR in discs with
uniform kinematics, although the samples considered are rather small
and also prone to biases \citep{Flores:06,Puech:08}.

Although semi-analytical models of galaxy formation predict a redshift
evolution of the TFR in rough agreement with the observations,
important pieces in the modelling of star formation are still missing.
For instance, the models are not capable of reproducing the evolution
of the optical TFR with redshift, and predict too small disc scale
lengths \citep[see, e.g.][]{Ton:11}, reflecting the shortcomings of
the prescriptions, e.g., driving the fuelling and quenching of star
formation, or the dynamical evolution of the galaxy and the
redistribution of angular momentum during the collapse of the baryons
towards the centres of dark matter halos. In this paper, we apply a
simple phenomenological approach to gain insight on the TFR from the
point of view of the underlying stellar populations. The more
physically motivated version of the TFR, namely the correlation
between circular speed and stellar mass \citep[e.g.][]{BdJ:01}, is an
important indicator of disc galaxy formation. Comparisons with
evolution models allow us to use stellar mass instead of luminosity,
reducing the highly variable changes in optical luminosity caused by
the presence of young stars. The focus of this paper is to deconstruct
the stellar mass TFR by exploring the age distribution of the stellar
populations. In \citet{FSBZ:04} we already showed that the stellar
ages of disc galaxies at moderate redshift were a strong function of
galaxy mass, possibly explained by a significant change in the star
formation efficiency, with a strong increase at a circular speed
$V_{\rm MAX}>140$\,km\,s$^{-1}$.  The mass-dependence of the stellar
population properties of galaxies is a well-known phenomenon in the
local universe \citep[e.g.][]{Kauff:03}. Age indicators, like the
strength of the 4000 \AA\ break \citep{Bruzual:83}, increase towards
higher stellar masses, implying that high-mass galaxies formed stars
with higher efficiency than low-mass ones in the cosmic past
\citep{PPG:08}.  The connection between the mass of a galaxy and its
stellar populations is also deeply intertwined with the evolution of
the processes that drive galaxy formation over cosmic
timescales. Overall, the main site of star formation shifts from
high-mass galaxies at high redshift to successively lower-mass
galaxies towards lower redshifts. This evolutionary trend is often
referred to as downsizing \citep[see,
  e.g.][]{Cowie:96,Kodama:04,Cimatti:06}.  Several feedback processes
have been proposed to explain such relationships, e.g.~supernova
feedback that regulates star formation in low-mass galaxies
\citep[e.g.][]{Gov:09}, or suppression of star formation by active
galactic nuclei in the center of high-mass galaxies
\citep[e.g.][]{Khalat:08}.  It is still an ongoing debate whether the
governing factor in such correlations is the stellar mass of a galaxy
or rather the mass of its host dark matter halo. In the latter case,
environment rather than mass would actually be the key factor that
shapes the stellar populations.  However, studies based on very large
samples from the Sloan Digital Sky Survey
\citep[e.g.][]{vdBosch:08,Rogers:10} present evidence that, at least
at the present cosmic epoch, galaxy properties such as optical colour
or morphological concentration are tightly correlated with stellar
mass but only weakly with halo mass.

In this paper, we revisit the evolution of disc galaxies by studying a
sample at moderate redshift (z$\sim$0.5), with the aim of probing the
evolution of the TFR and other scaling relations with respect to
stellar ages. The outline of the paper is as follows: We present the
photometric data from the FDF and WHDF samples (Section 2), and the
phenomenological model to describe the star formation histories,
including a comparison with more standard models of galactic chemical
enrichment (Section 3), followed by a discussion of the main results
(Section 4), along with our conclusions (Section 5).  Throughout this
paper, a standard $\Lambda$CDM cosmology is assumed, with
$\Omega_m=0.3$ and $H_0=70\,{\rm km\,s}^{-1}{\rm Mpc}^{-1}$.

%%%%%%%%%%%%%%%%%%%%%%%%%%%%%%%%%%%%%%%%%%%%%%%%
%%%%%%%%%%%%%%%%  Figure 1   %%%%%%%%%%%%%%%%%%%
%%%%%%%%%%%%%%%%%%%%%%%%%%%%%%%%%%%%%%%%%%%%%%%%
\begin{figure*}
  \begin{center}
    \includegraphics[width=15cm]{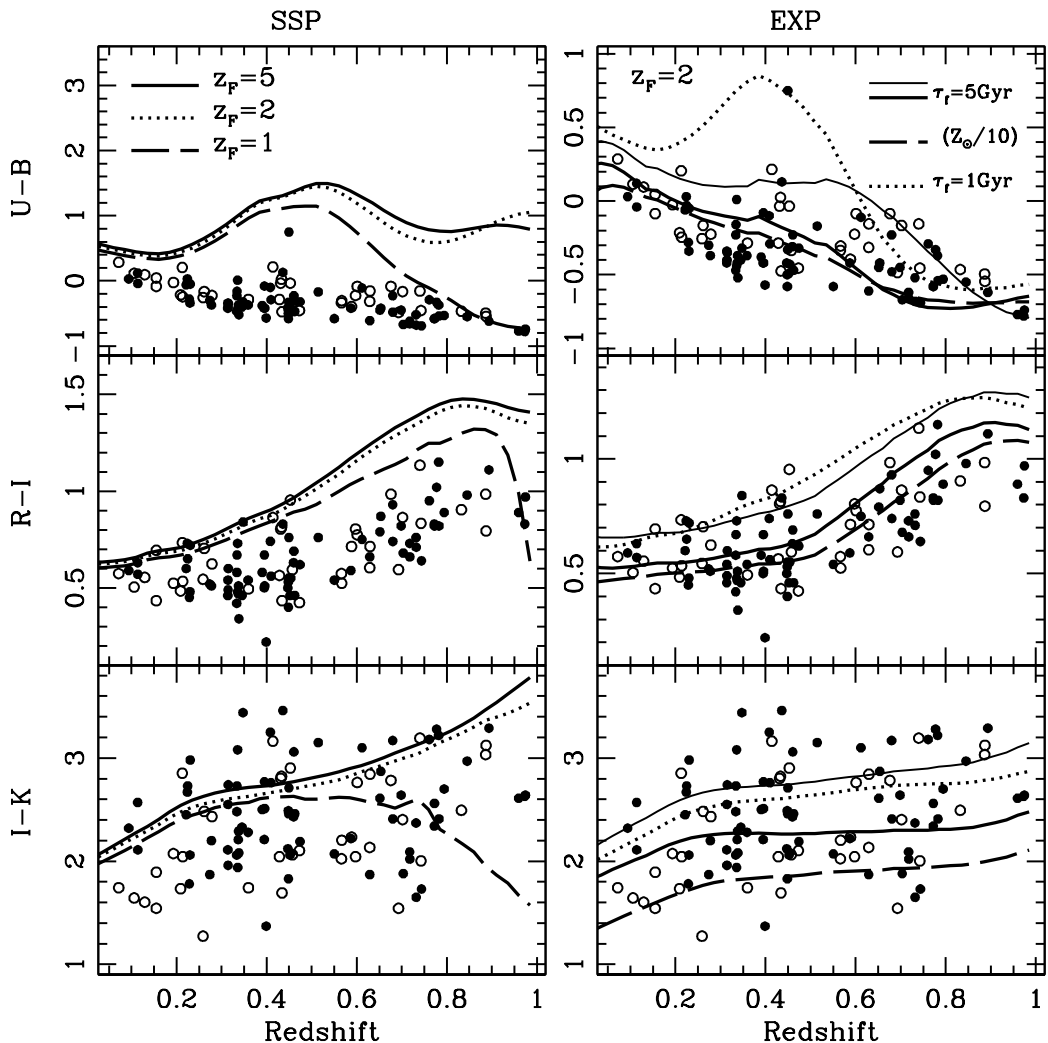}
  \end{center}
  \caption{Comparison of the photometry with simple stellar
    populations ({\sl left}) and models with exponentially decaying
    star formation histories at fixed metallicity ({\sl right}), all
    based on the \citet{BC03} population synthesis models. The thin
    solid line in the EXP panels corresponds to a colour excess of
    $E(B-V)=0.2$\,mag for a $\tau_f=5$\,Gyr model, at solar
    metallicity. Galaxies in the FDF (WHDF) are shown as solid (open)
    dots, where a small offset (always below $0.1$\,mag) is applied to
    the WHDF data to correct for the different response of the filters
    used (only for display purposes in this figure).
   \label{fig:photo}}
\end{figure*}
%%%%%%%%%%%%%%%%%%%%%%%%%%%%%%%%%%%%%%%%%%%%%%%%

%%%%%%%%%%%%%%%%%%%%%%%%%%%%%%%%%%%%%%%%%%%%%%%%
\section{The Sample}

A sample of field disc galaxies is selected from the multi-band
imaging of the FORS Deep Field \cite[FDF, ][]{fdf} and William
Herschel Deep Field \cite[WHDF, ][]{whdf} surveys. This sample was
originally selected for the analysis of disc kinematics in the
redshift range $0.1<z<1.0$, using follow-up spectroscopic data taken
with the FORS camera at the Very Large Telescope \citep[see,
  e.g.,][]{Bohm:04,Bohm:07}. In multi-object spectroscopy mode, a slit
was placed along the apparent major axis of each target to extract a
rotation curve, i.e.~the rotation velocity as a function of radius.
{\sl Hubble Space Telescope} imaging with the Advanced Camera for
Surveys (ACS) was used to determine parameters such as position angle,
disc inclination or scale length, with the use of the GALFIT package
\citep{galfit}. We derived the maximum rotation velocity, $V_{\rm
  MAX}$, taking into account disc inclination, luminosity profile, the
angle between the slit and the apparent major axis ($<$15$^{\rm o}$
for all galaxies), the influence of the slit width, and the seeing.
In order to obtain robust estimates of the maximum circular speed, we
discarded objects with a) disturbed kinematics, b) rotation curves
that did not reach the turnover region (i.e. apparent solid-body
rotation) or c) a too low signal-to-noise ratio. This resulted in a
final sample of $124$ galaxies, with a median uncertainty of
$\Delta\log V_{\rm MAX}=0.087$\,dex.

Photometry of our sample is available from the FDF and WHDF survey
data. The FDF photometry has been acquired with the Very Large
Telescope in the $U$-, $B$-, $g$-, $R$- and $I$-bands, and the New
Technology Telescope in the $J$- and $K_s$-bands. The WHDF photometry
was carried out at the William Herschel Telescope in the $U$-, $B$-,
$R$-, and $I$-bands and the Calar Alto 3.5m telescope in the $H$- and
$K$-bands.  Fixed aperture photometry was performed after convolving
all frames to a common PSF with FWHM 1 (1.5)\,arcsec for FDF (WHDF)
galaxies. The aperture diameter used for the photometry is 2 (3)\,arcsec
for FDF (WHDF) sources. The final magnitudes were corrected for Galactic extinction
following \citet{Card89}, adopting a reddening $E(B-V)=0.018$\,mag and
$E(B-V)=0.030$\,mag towards the positions of the FDF and WHDF,
respectively \citep[based on ][]{Schlegel98}. A number of galaxies
with available $V_{\rm MAX}$ values were not observed in all filters
and thus rejected, resulting in a final sample of $108$ discs ($73$ in
FDF and $35$ in WHDF), between $z=0.07$ and $z=0.97$ at a median
redshift of $\langle z \rangle = 0.45$.

Star formation rates (SFRs) were estimated from [O\,{\small II}]
equivalent widths, following \citet{Ken92} (other emission lines, such
as H$\beta$ or [O\,{\small III}] are sometimes used for the
determination of the rotation curves, depending on the wavelength
range probed by a given spectrum).  Note that H$\alpha$ is not
available within the observed wavelength range to derive SFRs given
the redshift distribution of the sample. In order to obtain a
consistent estimate of SFRs, we discarded $43$
objects that do not contain [O\,{\small II}] either because the
redshift is too low, or because the slit position on the CCD resulted
in a high starting wavelength, shifting [O\,{\small II}] bluewards
of the covered wavelength range. The final sample with SFRs therefore
comprises $65$ galaxies.

Figure~\ref{fig:photo} shows a comparison of some of the available
colours with respect to two sets of synthetic models from
\citet{BC03}: simple stellar populations (i.e. single age and
metallicity, {\sl left}) or composite models with an exponentially
decaying star formation history, at fixed metallicity
(i.e. $\tau$-models, {\sl right}). Different lines correspond to
choices of stellar age in SSP models (given by the formation
redshift), or exponential timescale and metallicity for the $\tau$
models, as labelled.  The combination of colours across a wide range
of wavelengths allows us to constrain the ages and metallicities: the
optical colours ({\sl top}) are most sensitive to age, whereas the NIR
colours have an increased dependence with respect to metallicity. All
these models are dustless, except for the thin solid line in the
$\tau$ models, which corresponds to an intrinsic reddening $E(B-V)=0.2$
\citep[following the][extinction law]{Card89}. The comparison
shows that the reddening values stay below $E(B-V)\simlt 0.5$\,mag. 
The figure also illustrates the need to choose composite models to
explain the multi-band photometric data, as expected from the complex
star formation histories of disc galaxies.

%%%%%%%%%%%%%%%%%%%%%%%%%%%%%%%%%%%%%%%%%%%%%%%%
%%%%%%%%%%%%%%%%  Figure 2   %%%%%%%%%%%%%%%%%%%
%%%%%%%%%%%%%%%%%%%%%%%%%%%%%%%%%%%%%%%%%%%%%%%%
\begin{figure}
  \begin{center}
    \includegraphics[width=3.3in]{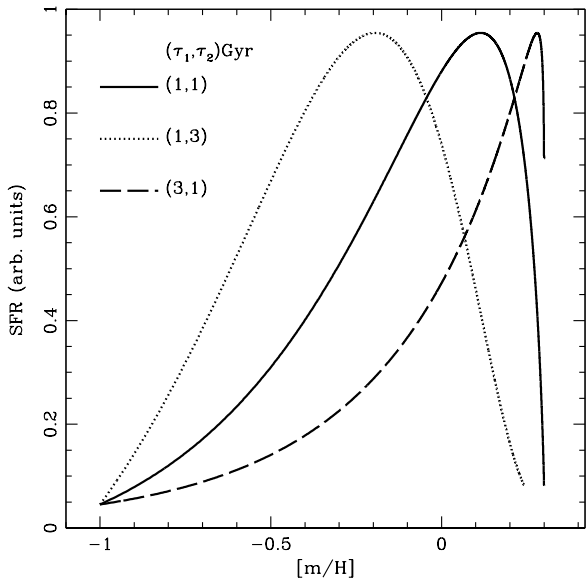}
  \end{center}
  \caption{The model adopted to track the star formation histories of
    disc galaxies includes two different timescales: $\tau_1$ tracks
    the star formation rate -- which evolves as a delayed
    exponential. $\tau_2$ controls the enrichment rate. This figure
    shows the difference in the time evolution of the star formation
    rate and the metallicity ([m/H]) for three cases, as labelled. An
    additional parameter (formation redshift) controls the average age
    of the distribution.
   \label{fig:sfh}}
\end{figure}
%%%%%%%%%%%%%%%%%%%%%%%%%%%%%%%%%%%%%%%%%%%%%%%%

%%%%%%%%%%%%%%%%%%%%%%%%%%%%%%%%%%%%%%%%%%%%%%%%
\section{Modelling Star Formation Histories}

\begin{figure}
  \begin{center}
    \includegraphics[width=3.5in]{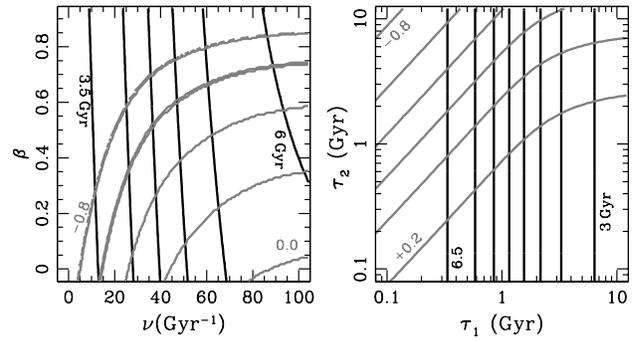}
  \end{center}
  \caption{Comparison of the model used in this paper with a more
    generic chemical enrichment model that includes parameters to
    describe the star formation efficiency ($\nu$) and the fraction of
    gas ejected in outflows ($\beta$) \citep[see e.g.][for
      details]{FS:01}.  In order to simplify the comparison, we assume
    the same formation epoch in both cases ($z_{\rm FOR}=5$). The
    panel on the left shows contours of mass-weighted average age
    (black) and metallicity (grey), with respect to star formation
    efficiency and gas outflow fraction. The black (grey) labels
    indicate the extrema in age (metallicity). The panel on the right
    shows similar contours for the model used here, exploring a range
    of values of star formation timescale ($\tau_1$) and enrichment
    timescale ($\tau_2$).
   \label{fig:chem}}
\end{figure}

We apply a phenomenological one-zone model that describes the star
formation history (SFH) and chemical enrichment of a galaxy with three
parameters. The model is constrained by the broadband photometric data
presented above. A simple model allows us to probe a large volume of
parameter space, and therefore of SFHs, giving robust constraints
about the mechanisms contributing to the build-up of the stellar
populations. The star formation history is assumed to start at an
epoch given by a formation redshift (z$_{\rm FOR}$), with a star
formation rate modelled either by a delayed exponential: \be
\psi(t)\propto [t-t(z_{\rm FOR})]e^{-[t-t(z_{\rm FOR})]/\tau_1}, \ee
\noindent
or by a standard exponential (i.e. a $\tau$-model):
\be 
\psi(t)\propto e^{-[t-t(z_{\rm FOR})]/\tau_1}, 
\ee
\noindent
and a metallicity trend driven by the growth in stellar mass, although
with an independent timescale, to take into account the effects of
both gas outflows and star formation efficiency: 
\be 
Z(t)\equiv Z_1 + Z_2\Big[ 1-\exp^{-[t-t(z_{\rm FOR})]/\tau_2}\Big].  
\ee
\noindent
The extrema in metallicity are $Z_1=Z_\odot/10$ and $Z_2=2Z_\odot$.
$\tau_1$ represents the timescale for the formation of the stellar
component, whereas $\tau_2$ drives the metal enrichment. In this
  paper, we consider as free parameters $\tau_1$, $\tau_2$ and the
  formation epoch, z$_{\rm FOR}$, allowing for a wide range of star
  formation histories with a decoupling between star formation, gas
infall and outflows. As an example, Fig.~\ref{fig:sfh} shows the
relationship between star formation rate (assuming a delayed
exponential law) and metallicity for three models. Note that the
dotted line, corresponding to a balance between enrichment and infall,
gives a metallicity distribution similar to the closed box model,
producing an excess of stars at low metallicity. This so-called
G-dwarf problem \citep[see, e.g.,][]{pagel} -- is solved in this
framework by a fast enrichment (short $\tau_2$) along with an extended
period of infall (long $\tau_1$). To illustrate the validity of these
models, we compare the age and metallicity distributions with a more
detailed analysis based on standard prescriptions of chemical
enrichment. These more detailed models describe the build up of
  the stellar populations and their metallicities by a process of gas
  infall/outflow -- where the infall and outflow rates are controlled
  by free parameters -- along with a \cite{Schmidt:63} law to drive
  the transformation of gas into stars, and an additional parameter
  for the star formation efficiency that regulates the fraction of the
  gas component that is transformed into stars per unit time
  \citep[see][for details]{FS:01,FSBZ:04}. In Fig.~\ref{fig:chem},
contours of average stellar age (black lines) and metallicity (grey
lines) are shown with respect to star formation efficiency ($\nu$) and
outflow fraction ($\beta$) for the enrichment model ({\sl left}) or
with respect to the two timescales ($\tau_1$ and $\tau_2$) for the
phenomenological model used here ({\sl right}). The formation epoch is
fixed in both cases to $z_{\rm FOR}=5$. A mapping can be established
between star formation efficiency ($\nu$) and our $\tau_1$; and also
between outflow fraction ($\beta$) and our $\tau_2$. We emphasize that
the model presented here is meant to give a simple and clear
representation of the formation of the stellar populations in disc
galaxies.

%%%%%%%%%%%%%%%%%%%%%%%%%%%%%%%%%%%%%%%%%%%%%%%%
%%%%%%%%%%%%%%%%  Figure 3   %%%%%%%%%%%%%%%%%%%
%%%%%%%%%%%%%%%%%%%%%%%%%%%%%%%%%%%%%%%%%%%%%%%%
%%%%%%%%%%%%%%%%%%%%%%%%%%%%%%%%%%%%%%%%%%%%%%%%

%%%%%%%%%%%%%%%%%%%%%%%%%%%%%%%%%%%%%%%%%%%%%%%%
\section{Results}

The models presented in the previous section are explored over a large
range of parameters.  We also include the effects of dust as a
homogeneous screen. We describe this component as an additional fourth
free parameter, choosing the colour excess, $E(B-V)$, from the
\citet{Card89} extinction law. In order to obtain the most robust
estimates of the best fit {\sl and uncertainty}, we opt for a
comprehensive search of parameter space. This is doable given the
small number of parameters describing the star formation history.
Furthermore, a parameter search algorithm has a very simple
parallelisation scheme, so that we can run it efficiently on many
processors. For a choice of the four parameters, the resulting star
formation history is combined with the synthetic populations of
\citet{BC03} to derive the six (five) broadband photometric colours
available from the FDF (WHDF) data, generating a likelihood (via a
standard $\chi^2$ statistic), that defines a probability distribution
function for the three parameters under consideration. Out of the two
families of models -- corresponding to either a delayed exponential,
or a $\tau$ model -- we select for each galaxy the one that gives the
lowest value of the $\chi^2$. Although discriminating between these
two prescriptions for the star formation history is beyond the
capabilities of broadband photometry, we note that there is a slight
preference towards the delayed exponential model, especially at low
stellar mass. Roughly in 70\% of the sample a delayed exponential
model is preferred over a $\tau$ model. However, the trends in the
average ages and metallicities are not affected by the choice. We note
there is a 1:1 correspondence between the predicted age in both cases
for populations younger than $\sim$3\,Gyr, whereas delayed exponential
models give ages $\simlt 1$\,Gyr younger than the $\tau$ models for
older populations. The trend with metallicity is similar, with a 1:1
correspondence at solar/supersolar metallicities and offsets
$\simlt$0.2\,dex more metal rich in the delayed exponential models.
We derive from the likelihood the average mass-weighted age and
metallicity, along with the uncertainties, quoted throughout this
paper at the 69\% level.  For each galaxy, we run two grids of
$32^3\times 16$ models (one for each choice of star formation rate),
uniformly scaled over the following range of the parameters:
\begin{equation*}
\begin{aligned}
 1\leq &z_{\rm FOR}\leq 5\\
-1\leq &\log(\tau_1/{\rm Gyr})\leq +1\\
-1\leq &\log(\tau_2/{\rm Gyr})\leq +1\\
0\leq & E(B-V)\leq 0.5\\
\end{aligned}
\end{equation*}

%%%%%%%%%%%%%%%%%%%%%%%%%%%%%%%%%%%%%%%%%%%%%%%%
%%%%%%%%%%%%%%%%  Figure 4   %%%%%%%%%%%%%%%%%%%
%%%%%%%%%%%%%%%%%%%%%%%%%%%%%%%%%%%%%%%%%%%%%%%%
\begin{figure}
  \begin{center}
    \includegraphics[width=3.3in]{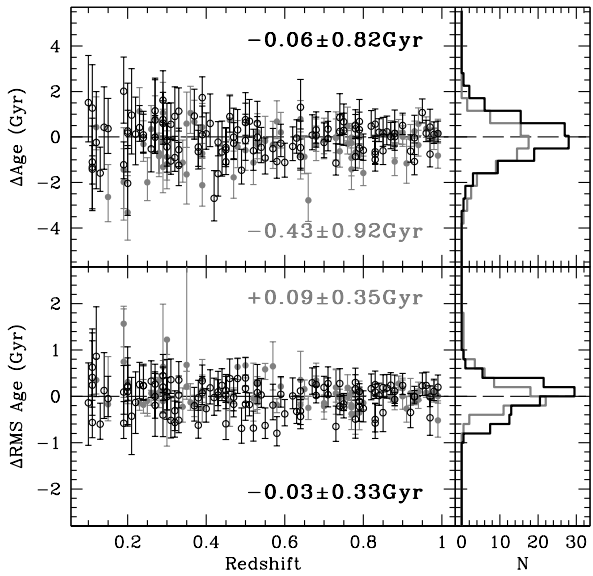}
  \end{center}
  \caption{Recovery of average (top)
    and RMS of the age distribution (bottom) from $200$ simulations
    spanning a similar range of redshift and SNR as the original
    data. The vertical axis in each panel represents input minus
    recovered value.  The black (grey) data points use the metallicity
    prior from the \citet{gallazzi:05} relation, assuming a stellar
    mass of $10^{11}$\,M$_\odot$ ($10^9$\,M$_\odot$).  Error bars are
    the 69\,\% confidence levels from the analysis, and the histogram
    of the distribution is shown in the rightmost panels. The values
    of the difference in average age and RMS of each case is indicated
    with the same colour coding.
\label{fig:sims}}
\end{figure}
%%%%%%%%%%%%%%%%%%%%%%%%%%%%%%%%%%%%%%%%%%%%%%%%

In addition, we impose a prior on the average stellar metallicity,
using the local mass-metallicity relation of \citet{gallazzi:05}. We
note that this prior is rather mild, given the large scatter of the
distribution at fixed mass ($\Delta[Z/H]\!\sim\!1$\,dex at $\log
M_s/M_\odot=10$). The reason for applying this prior is based on the
lack of a large number of independent metallicities in the synthetic
models, along with the fact that broadband photometry alone -- only
extending out to rest-frame data in the $H$ band, given the redshift
range -- cannot impose strong constraints on the metallicities of
unresolved stellar populations.  This prior helps to ensure a correct
age-metallicity discrimination in the analysis. The error
bars on the estimates of stellar metallicity are around
$\Delta[Z/H]\!\sim\!0.3$\,dex and the average age uncertainties
$\Delta\langle{\rm Age}\rangle/\langle{\rm Age}\rangle=0.4$. A
comparison between the ages derived with and without the prior on
metallicity gives a difference $\langle{\rm Age}\rangle_{\rm PRIOR} -
\langle{\rm Age}\rangle_{\rm NO\ PRIOR} = 0.14\pm 0.61$\,Gyr, without
any systematic effect with respect to galaxy mass.

%%%%%%%%%%%%%%%%%%%%%%%%%%%%%%%%%%%%%%%%%%%%%%%%
%%%%%%%%%%%%%%%%  Figure 5   %%%%%%%%%%%%%%%%%%%
%%%%%%%%%%%%%%%%%%%%%%%%%%%%%%%%%%%%%%%%%%%%%%%%
\begin{figure}
  \begin{center}
    \includegraphics[width=3.3in]{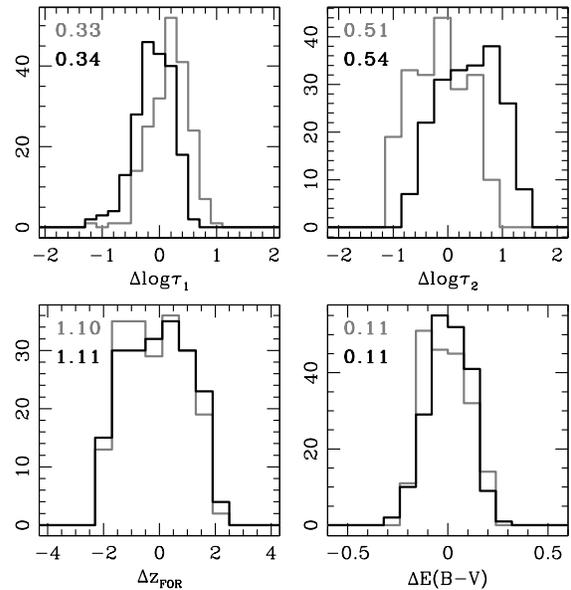}
  \end{center}
  \caption{Recovery of model parameters from $200$ simulations
    spanning a similar range of redshift and SNR as the original
    data. In each case $\Delta$ is defined as the difference between
    the input and the recovered parameter.  The black (grey) histograms
    correspond to the application of the prior regarding the
    mass-metallicity relationship of \citet{gallazzi:05} assuming
    galaxies with stellar mass of $10^9$\,M$_\odot$
    ($10^{11}$\,M$_\odot$), to illustrate the effect of such a prior
    on the recovery of the model parameters (see text for
    details). The numbers in each panel show the RMS of the
    distributions for each choice of stellar mass, with the same
    colour coding.
\label{fig:sims_pars}}
\end{figure}
%%%%%%%%%%%%%%%%%%%%%%%%%%%%%%%%%%%%%%%%%%%%%%%%

%%%%%%%%%%%%%%%%%%%%%%%%%%%%%%%%%%%%%%%%%%%%%%%%
%%%%%%%%%%%%%%%%  Figure 6   %%%%%%%%%%%%%%%%%%%
%%%%%%%%%%%%%%%%%%%%%%%%%%%%%%%%%%%%%%%%%%%%%%%%
\begin{figure}
  \begin{center}
    \includegraphics[width=3.3in]{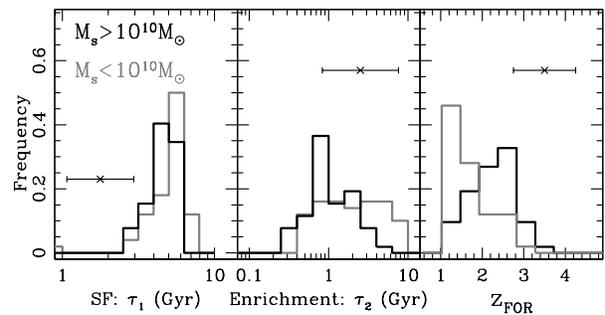}
  \end{center}
  \caption{The distribution of the parameters $\tau_1$ (star formation
    timescale), $\tau_2$ (enrichment timescale), and z$_{\rm FOR}$
    (formation epoch) is shown for our sample of FDF+WHDF disc
    galaxies, split with respect to stellar mass, as labelled. A
    typical 1\,$\sigma$ error bar is included in all panels.
  \label{fig:t1t2}}
\end{figure}
%%%%%%%%%%%%%%%%%%%%%%%%%%%%%%%%%%%%%%%%%%%%%%%%

%%%%%%%%%%%%%%%%%%%%%%%%%%%%%%%%%%%%%%%%%%%%%%%%
%%%%%%%%%%%%%%%%  Figure 7   %%%%%%%%%%%%%%%%%%%
%%%%%%%%%%%%%%%%%%%%%%%%%%%%%%%%%%%%%%%%%%%%%%%%
\begin{figure*}
  \begin{center}
    \includegraphics[width=15cm]{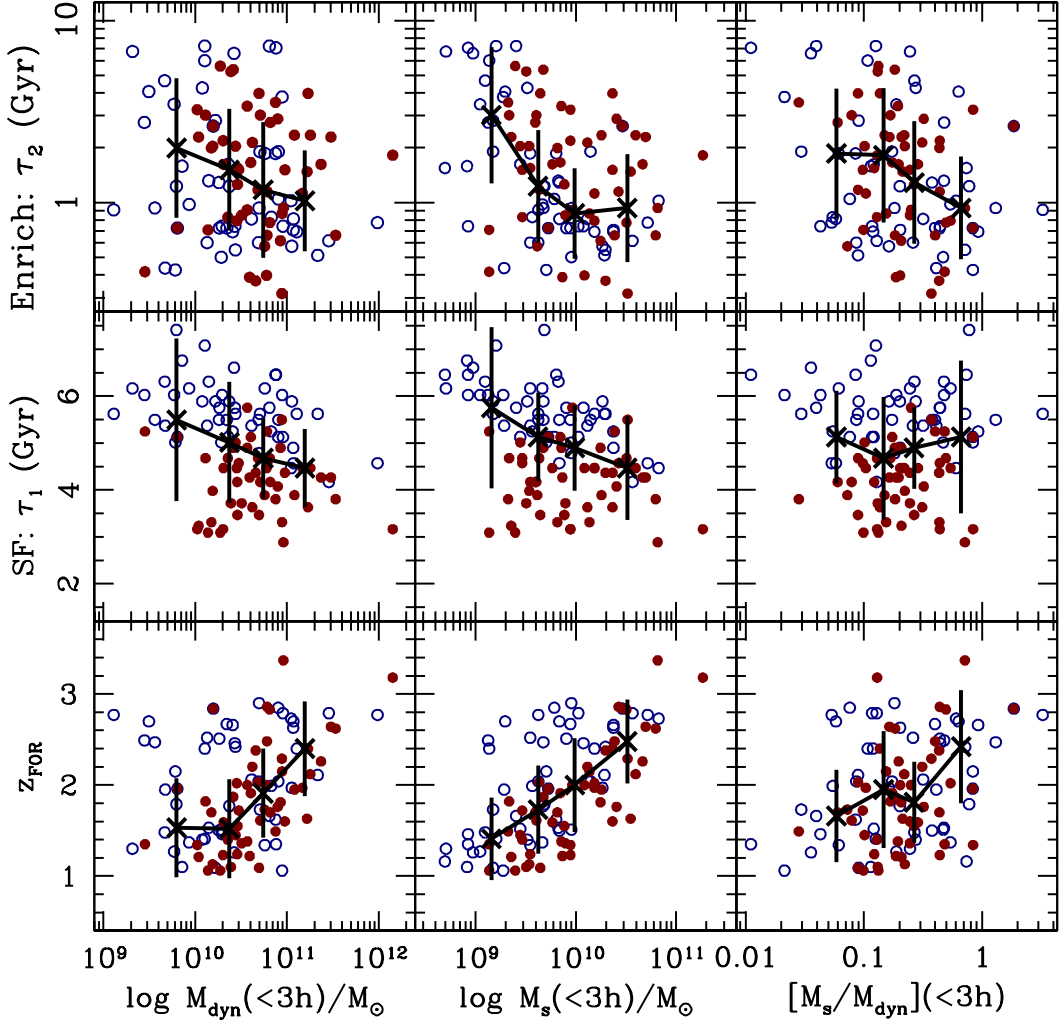}
  \end{center}
  \caption{The constraints on the parameters of the models are shown
    as a function of total mass ({\sl left}), stellar mass ({\sl centre}) and
    stellar-to-total mass fraction ({\sl right}). Individual values are
    shown as dots, whereas the black crosses represent the median
    values after binning (with equal number of galaxies per bin),
    including the RMS scatter per bin as an error bar. The red solid
    (blue open) dots correspond to $z\geq 0.45$ ($<0.45$) (i.e. split
    at the median redshift of the sample).
\label{fig:cxp}}
\end{figure*}
%%%%%%%%%%%%%%%%%%%%%%%%%%%%%%%%%%%%%%%%%%%%%%%%

In order to confirm this point, we performed 200 simulations with the
same redshift and signal-to-noise distribution as the original data,
comparing the age and RMS of the age distribution. Fig.~\ref{fig:sims}
shows the difference between the input and the recovered values for the
average age ({\sl top}) and the RMS of the age distribution ({\sl
  bottom}).  The simulations assume two different cases of stellar
mass for the application of the prior: $10^{11}$\,M$_\odot$ (black)
and $10^9$\,M$_\odot$ (grey). In either case, the recovered values do
not show any systematic offset, with an accuracy of $\sim\!0.9$\,Gyr in
average age and $\sim\!0.3$\,Gyr in the RMS of the age distribution.
In Fig.~\ref{fig:sims_pars} a similar comparison is made, in this case
between the model parameters. The numbers in each panel give the RMS
of the distributions of the (input~$-$~output) values of the parameters
for the simulations. Note that there is no systematic change in the
values of the parameters, except for $\tau_2$. In this case, as expected,
the prior biases the values of metallicity towards lower 
metallicity for the $10^9$\,M$_\odot$ (grey) case, and towards higher
metallicity at $10^{11}$\,M$_\odot$ (black).  We note that our
approach is quite conservative, as the prior for the simulations is
blindly applied to all galaxies, irrespective of their input value of
metallicity.

\subsection{Model parameters}

Hereafter, we present the results for the FDF/WHDF data.
Fig.~\ref{fig:t1t2} shows the distribution of the two formation
timescales and the formation epoch. The sample is split with respect
to stellar mass, at the position of the median
($\sim\!10^{10}$M$_\odot$).  For reference, the median uncertainty (at
the 68\% level) is shown in each panel as a horizontal error bar. Note
the difference between the distribution of low- and high-mass discs,
especially in the enrichment timescale, $\tau_2$ and the formation
epoch. The majority of low-mass discs (grey histograms) are fitted by
longer timescales both in star formation ($\tau_1$) and enrichment
($\tau_2$), along with later formation times.  A large fraction (56\%)
of the massive galaxies (M$_s>10^{10}$M$_\odot$) are better explained
by very short enrichment timescales ($\tau_2\simlt 1$\,Gyr), in
contrast with 28\% for the low-mass subsample. Low values of {\sl
  both} $\tau_1$ and $\tau_2$ would represent the efficient and fast
buildup of the stellar populations expected in early-type galaxies
\citep[see, e.g.,][]{dlR11}.  In contrast, massive discs have more
extended star formation timescales (reflecting a lower star formation
efficiency or an extended infall), and short enrichment timescales
(possibly caused by a lack of metal-rich outflows). The difference in
the error bars for $\tau_1$ and $\tau_2$ is caused by the higher
accuracy on the age estimates, with respect to metallicity.

%%%%%%%%%%%%%%%%%%%%%%%%%%%%%%%%%%%%%%
%%%%%%%%%%   TABLE 1   %%%%%%%%%%%%%%%
%%%%%%%%%%%%%%%%%%%%%%%%%%%%%%%%%%%%%%
\begin{table*}
\begin{center}
\caption{Non-parametric Spearman's Rank Correlation values for the
  four free parameters of the models (see Figs.~\ref{fig:cxp} and
  \ref{fig:dust}). The numbers in brackets give the 90th percentile of
  an ensemble of 1000 realisations where the data points were randomly
  shuffled 1000 times in each of the realisations.}
\label{tab:spearman}
\begin{tabular}{cccccc}
\hline
X & Redshift & \multicolumn{4}{c}{Y}\\
\cline{3-6}
  & Range    & $\log\tau_1$ & $\log\tau_2$ & z$_{\rm FOR}$ & E$_{\rm B-V}$\\
\hline
$\log$M$_s$  &  All     & $-0.4403$  & $-0.3694$  & $+0.6627$  & $+0.7592$\\
             &          & $(0.1247)$ & $(0.1213)$ & $(0.1261)$ & $(0.1225)$\\
             & z$<0.45$ & $-0.5098$  & $-0.5928$  & $+0.6205$  & $+0.7518$\\
             &          & $(0.1970)$ & $(0.1762)$ & $(0.1869)$ & $(0.1802)$\\
             & z$>0.45$ & $-0.3540$  & $-0.1913$  & $+0.7980$  & $+0.8004$\\
             &          & $(0.1830)$ & $(0.1784)$ & $(0.1762)$ & $(0.1908)$\\
\hline          
$\log$M$_{\rm dyn}$ &  All & $-0.2839$ & $-0.1399$ & $+0.4566$  & $+0.4659$\\
             &          & $(0.1272)$ & $(0.1274)$ & $(0.1127)$ & $(0.1313)$\\
             & z$<0.45$ & $-0.2279$  & $-0.2770$  & $+0.2661$  & $+0.3523$\\
             &          & $(0.1840)$ & $(0.1897)$ & $(0.1762)$ & $(0.1721)$\\
             & z$>0.45$ & $-0.2416$  & $-0.0736$  & $+0.7630$  & $+0.5634$\\
             &          & $(0.1982)$ & $(0.1709)$ & $(0.1762)$ & $(0.1808)$\\
\hline
$\log($M$_s$/M$_{\rm dyn})$ & All  & $-0.1044$ & $-0.2317$ & $+0.2122$ & $+0.2716$\\
             &          & $(0.1199)$ & $(0.1291)$ & $(0.1210)$ & $(0.1300)$\\
             & z$<0.45$ & $-0.1989$  & $-0.2269$  & $+0.2664$  & $+0.3002$\\
             &          & $(0.1814)$ & $(0.1646)$ & $(0.1603)$ & $(0.1761)$\\
             & z$>0.45$ & $-0.0874$  & $-0.2846$  & $+0.2084$  & $+0.4036$\\ 
             &          & $(0.1761)$ & $(0.1896)$ & $(0.1671)$ & $(0.1783)$\\
\hline 
\end{tabular}\\
\end{center}
\end{table*}
%%%%%%%%%%%%%%%%%%%%%%%%%%%%%%%%%%%%%%%%%%%%%%%%%%%%%%%%%%%

Fig.~\ref{fig:cxp} shows the best fit parameters as a function of
physical properties: dynamical mass ({\sl left}), stellar mass ({\sl
  centre}), or stellar-to-dynamical mass fraction ({\sl right}). To
avoid model-dependent extrapolations for the dark matter distribution,
both dynamical and stellar masses are quoted within a fixed aperture
of $3$ exponential scale lengths (roughly where all galaxies in the
sample reach a flat circular velocity), i.e. the dynamical mass is
given as M$_{\rm dyn}\equiv 3V_{\rm MAX}^2r_d/G$, where $r_d$ is the
disc scale length.  The scale lengths are derived from surface
brightness fits to the ACS images \citep{Bohm:07}. The average
fractional error for the sample explored in this paper is 20\% in the
disc scale length. Including the 0.087\,dex error in $V_{\rm MAX}$
(propagated from uncertainties in inclination and the fit of the
rotation curve) results in an average error of $\Delta\log M_{\rm
  dyn}\sim 0.17$\,dex. The stellar mass uncertainty -- obtained from
the PDF derived by the large volume of parameter space explored by the
models -- is $\Delta \log M_s\simlt 0.07$\,dex. For a clearer
interpretation of the trends, the individual data points are arranged
into four bins. The binning is done at constant number of galaxies per
bin, and the median and dispersion (RMS) are shown as crosses, and
error bars, respectively. The individual data points are also coded
with respect to redshift (red solid: $z\geq 0.45$; open blue:
$z<0.45$). Note that the parameters are correlated with stellar mass,
with a clear downsizing trend with $z_{\rm FOR}$.

Tab.~\ref{tab:spearman} gives the non-parametric correlations
according to the Spearman test \citep[see, e.g.,][]{NR} for all four
free parameters explored by the models.  Values close to $+1$ ($-1$)
reflect a strong correlation (anticorrelation). The numbers in
brackets underneath each case correspond to the expected value for a
random distribution. It is obtained by a bootstrapping method, where
1000 realisations are created by randomly shuffling the data points
1000 times for each realisation. The quoted numbers (shown in
brackets) give the 90th percentile of this randomly generated
distribution of Spearman's correlation coefficients. Hence, numbers
smaller, in absolute number, to this figure would imply no
correlation. Notice the strongest correlation for the parameters shown
in Fig.~\ref{fig:cxp} is found between stellar mass and formation
epoch. The weak correlation between $\tau_2$ and M$_s$/M$_{\rm dyn}$
could be explained with the assumption that this fraction is an
indicator of efficiency or gas outflows, hence, at large stellar
masses, one should expect a fast, efficient buildup of metallicity
(i.e. a short $\tau_2$), however, we emphasize that the level of
correlation is at the border line, much weaker than any other trend
with stellar mass. Regarding redshift, the strongest trend is with
star formation/infall timescale ($\tau_1$, middle panels), reflecting
a more extended distribution of ages in the sample at low redshift,
mostly due to the wider range of ages because of lookback time.

%%%%%%%%%%%%%%%%%%%%%%%%%%%%%%%%%%%%%%%%%%%%%%%%
%%%%%%%%%%%%%%%%  Figure 8   %%%%%%%%%%%%%%%%%%%
%%%%%%%%%%%%%%%%%%%%%%%%%%%%%%%%%%%%%%%%%%%%%%%%
\begin{figure}
  \begin{center}
    \includegraphics[width=3.3in]{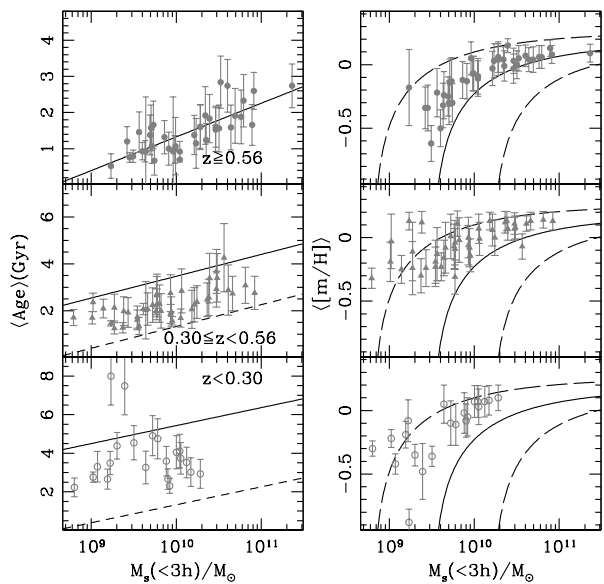}
  \end{center}
  \caption{Distribution of ages ({\sl left}) and metallicities ({\sl
      right}) of the FDF+WHDF disc galaxies, split with respect to
    redshift in three panels. The straight line on the top-left panel
    fits the data points in the high redshift subsample. This line is
    copied over, as a dashed line, in the other two redshift bins
    (middle- and bottom-left panels). We also show the effect of shifting
    this fit by the lookback time difference estimated at the average
    redshift within each bin, as solid lines in these
    two panels. On the right, the solid and dashed lines mark
    the average and $1\,\sigma$ uncertainties of the mass-metallicity
    relation of \citet{gallazzi:05} .
   \label{fig:tZ}}
\end{figure}
%%%%%%%%%%%%%%%%%%%%%%%%%%%%%%%%%%%%%%%%%%%%%%%%

%%%%%%%%%%%%%%%%%%%%%%%%%%%%%%%%%%%%%%%%%%%%%%%%
%%%%%%%%%%%%%%%%  Figure 9   %%%%%%%%%%%%%%%%%%%
%%%%%%%%%%%%%%%%%%%%%%%%%%%%%%%%%%%%%%%%%%%%%%%%
\begin{figure}
  \begin{center}
    \includegraphics[width=3.3in]{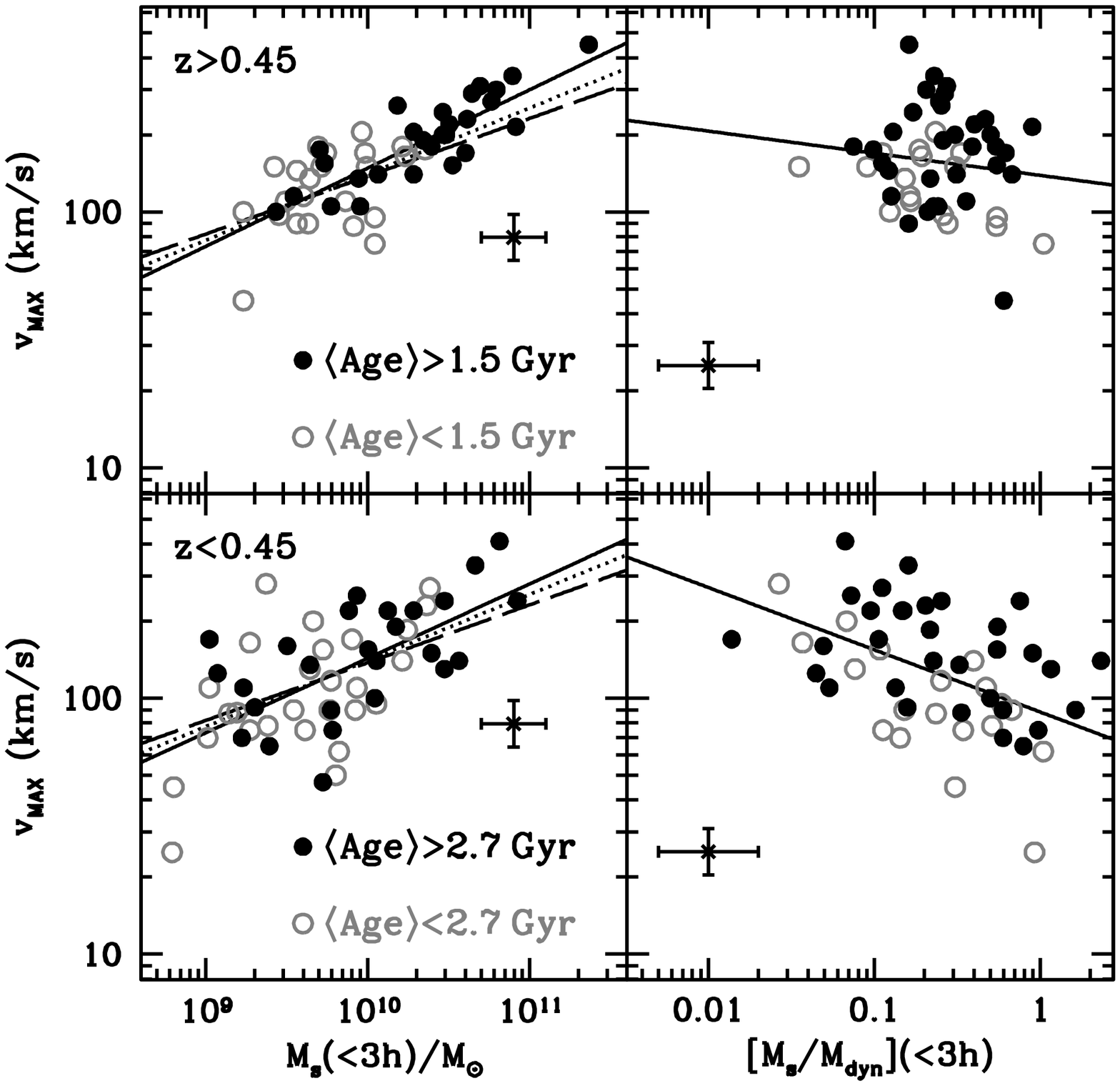}
  \end{center}
  \caption{Relationship between maximum rotation velocity and stellar
    mass ({\sl left}) or stellar-to-dynamical mass fraction ({\sl right}). The
    sample is split at the median with respect to stellar average age.
    Top and bottom panels show subsamples at high and low redshift,
    respectively (also split at the median of the distribution). The
    solid line in each panel is the best fit to the data within each
    redshift bin. Typical error bars are shown. The dashed line
    in the left panels show the local relationship from the $I$-band
    determined stellar masses \citep{BdJ:01}. The dotted line is
    the z$\sim$1 relationship from \citep{Miller:12}.
   \label{fig:TFz}}
\end{figure}
%%%%%%%%%%%%%%%%%%%%%%%%%%%%%%%%%%%%%%%%%%%%%%%%

The trends of average age and metallicity with stellar mass are shown
in Fig.~\ref{fig:tZ}, split into three redshift bins, that correspond
to uniform steps of $\sim\!2$\,Gyr in lookback times between
bins. Error bars are shown at the 68\% confidence level. In the
top-left panel of the figure, we fit the high redshift (z$\geq 0.56$)
subsample, creating copies of the same fit in the other two bins
(dashed lines on the middle-left and bottom left panels), also showing
a version shifted in age by the lookback time (solid lines).  When
taking into account the effect of lookback time, we find that the
average ages are $\sim 1$\,Gyr younger than the expectations from the
fit at high redshift, reflecting the fact that passive evolution does
not give a good representation to these galaxies.  Although this could
be interpreted as a lack of ''dynamical downsizing'', the sample is
not large enough to constrain a change in slope of the correlation
between age and stellar mass. As regards to metallicity, there is a
trend towards the metal rich envelope of the local mass-metallicity
relation.  Note the present analysis, based on broadband photometry
alone, mainly constrains stellar ages, whereas estimates of
metallicity carry a large error bar (typically
$\Delta[Z/H]\!\sim\!0.3$\,dex). Even though the local mass-metallicity
relation is used as a mild prior in the analysis, we stress that the
minimum values of $\chi^2$ are not significantly affected by the
prior, with $\sim\!90$\% of the sample having a reduced value
$\chi^2_r\simlt 2$.

\subsection{The evolution of the Tully-Fisher relation}

Fig.~\ref{fig:TFz} shows the properties of the underlying stellar
populations on a variation of the Tully-Fisher diagram
\citep{TF:77}. The sample is split at the median redshift (top/bottom
panels), and in each panel, black solid (grey open) circles represent
disc galaxies with an average age older (younger) than the mean value
within each redshift bin, as labelled. Typical ($1\,\sigma$) error
bars are shown as reference.  The panels on the left show a $V_{\rm
  MAX}$ vs stellar mass Tully-Fisher relation. The usual downsizing
relation between stellar mass and age is found. However, at fixed
stellar mass, no significant segregation with respect to $V_{\rm MAX}$
is evident. Old and young discs show a similar amount of scatter with
respect to the best linear fit (solid line). The slope and intercept
(see Tab.~\ref{tab:TFz}) are indistinguishable \citep[as also
  found by, e.g.][out to z$\sim$1.7]{Miller:11,Miller:12} with respect
to the redshift subsample chosen ($M_s\propto V_{\rm MAX}^{3.5}$).
The panels on the right show the maximum circular speed with respect
to stellar-to-dynamical mass fraction, with a clear trend towards a
faster $V_{\rm MAX}$ for disc galaxies with a lower stellar mass
fraction, as expected, since the stellar-to-total mass ratio decreases
with increasing galaxy mass \citep[see, e.g.,][]{moster10}.  At fixed
mass fraction, the younger galaxies have slower rotation speeds.  One
can assume that the stellar-to-dynamical mass fraction controls the
level of ``baryon feedback'' -- mainly related to star formation in
this sample. A tentative interpretation of the segregation with
respect to age, at fixed $M_s/M_{\rm dyn}$, would be that older
galaxies are preferentially assembled in higher density haloes,
therefore having a higher circular speed.

%%%%%%%%%%%%%%%%%%%%%%%%%%%%%%%%%%%%%%%%%%%%%%%%%%%%%
\subsection{The contribution of dust}

\begin{figure}
  \begin{center}
    \includegraphics[width=8.5cm]{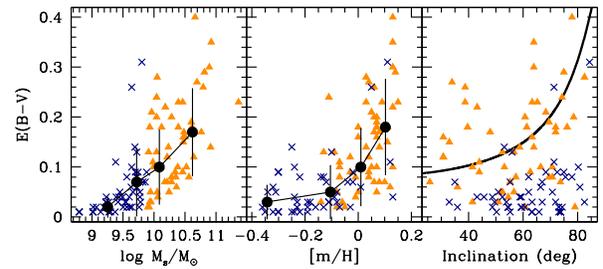}
  \end{center}
  \caption{Model constraints on the colour excess, with respect to
    stellar mass, metallicity or inclination. The orange triangles
    (grey crosses) correspond to stellar masses $\log
    M_s/M_\odot>10$ ($<10$).  The big solid dots with error bars in
    the left and middle panels give the median and RMS scatter of the
    complete sample, binned at fixed number of galaxies per bin. The
    rightmost panel includes a prescription from \citet{tf85}, with
    $f=0.2$ and $\tau=0.5$ (solid line).
   \label{fig:dust}}
\end{figure}

Fig.~\ref{fig:dust} shows the trend of dust reddening with respect to
stellar mass, average metallicity or inclination.  The sample is split
at the median value of the stellar mass ($\sim\!10^{10}$\,M$_\odot$),
with the orange triangles (blue crosses) corresponding to stellar
masses above (below) this cut. The big solid dots with error bars in
the left and middle panels give the median and RMS scatter of the
complete sample, binned at fixed number of galaxies per bin.  A
prescription from \citet{tf85} is included for reference, with $f=0.2$
and $\tau=0.5$ (solid line).  A strong trend is found with respect to
mass and metallicity, in good agreement with the observations of local
discs \citep{Tully:98}. The last column of Tab.~\ref{tab:spearman}
indicates that stellar mass, and not dynamical mass, is the main mass
observable correlated with colour excess.  More subtle is the excess
of more attenuated galaxies at higher inclinations for the more
massive galaxies. Notice that our modelling does not impose any prior
constraint on the colour excess, considering all values of $E(B-V)$ in
the analysis. The formalism of \citet{tf85} is applicable for the
massive disc subsample, whereas lower mass discs are not affected so
strongly with respect to disc inclination.

%%%%%%%%%%%%%%%%%%%%%%%%%%%%%%%%%%%%%%
%%%%%%%%%%   TABLE 2   %%%%%%%%%%%%%%%
%%%%%%%%%%%%%%%%%%%%%%%%%%%%%%%%%%%%%%
\begin{table}
\begin{center}
\caption{Slopes of the stellar mass-circular velocity Tully-Fisher
  relation (see Fig.~\ref{fig:TFz}). M$_{s,100}$ is the best fit
  stellar mass at $V_{\rm MAX}=100$\,km\,s$^{-1}$ Error bars quoted at
  the 1$\sigma$ level.}
\label{tab:TFz}
\begin{tabular}{ccc}
\hline
\multicolumn{3}{c}{M$_s=$M$_{s,100}(V_{\rm MAX}/V_{100})^\alpha$}\\
\hline
Redshift & $\alpha$ & $\log$ M$_{s,100}$\\
Range    &   & $/$M$_\odot$\\
\hline
 All      & $3.425\pm 0.447$ & $9.229\pm 0.097$\\
 z$<0.45$ & $3.675\pm 0.776$ & $9.236\pm 0.124$\\
 z$>0.45$ & $3.423\pm 0.724$ & $9.203\pm 0.184$\\
\hline
\end{tabular}
\end{center}
\end{table}
%%%%%%%%%%%%%%%%%%%%%%%%%%%%%%%%%%%%%%%%%%%%%%%%%%%%%%%%%%%

\subsection{Star formation rates (SFRs)}

\begin{figure}
  \begin{center}
    \includegraphics[width=3.3in]{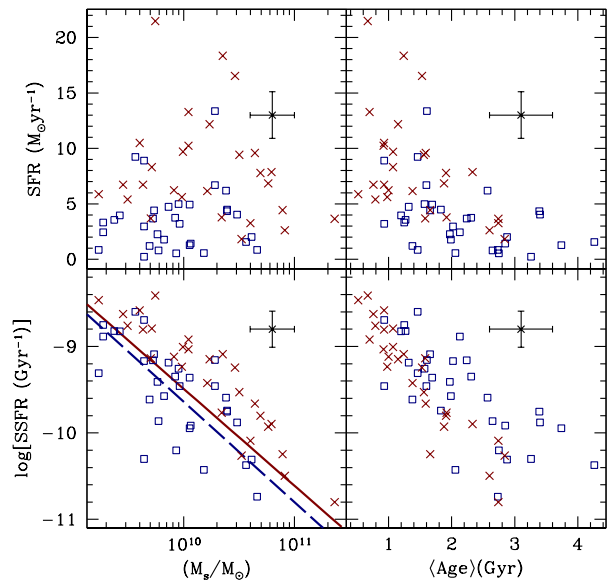}
  \end{center}
  \caption{Correlation of star formation rate (SFR, {\sl top}) and
    specific star formation rate (SSFR, {\sl bottom}) with respect to
    stellar mass ({\sl left}) or average age of the stellar
    populations ({\sl right}). The data are split at the median
    redshift of the subsample where SFR information is available (red
    crosses: z$>0.60$; blue open squares z$<0.60$). The blue dashed and red solid lines
    correspond to the ``main sequence'' of star forming galaxies at
    $z\sim 0.6$ and $z\sim 0.9$, respectively \citep{Brinch00}.
\label{fig:gas}}
\end{figure}

Ongoing star formation rates can be derived from emission line
luminosities.  In this paper, following \citet{Ken92}, we derive
  SFRs from the equivalent widths of [O\,{\small II}], and the
  $B$-band luminosity, namely:
\[
{\rm SFR} ({\rm M}_\odot\,\rm {yr}^{-1})\approx 2.7 \times 10^{-12} \frac{L_B}{L_B(\odot)}
EW([OII]) E(H_\alpha),
\]
where $E(H_\alpha)$ is the dust extinction correction factor at
  the wavelength of H$_\alpha$.  Fig.~\ref{fig:gas} presents the SFRs
({\sl top}) and specific SFRs (SSFR, defined as the SFR per unit
stellar mass, {\sl bottom}) with respect to either stellar mass ({\sl
  left}) or average age ({\sl right}).  The sample is split with
respect to redshift at the median value of the subsample comprising
$65$ galaxies where SFRs could be determined ($z_M=0.60$), with 
  blue open squares (red crosses) corresponding  to low (high)
redshift. Only a subset of the full sample is presented here because
the derivation of SFRs require estimates of [O\,{\small II}]
luminosities.  Characteristic error bars, at the $1\,\sigma$ level are
shown for reference. A strong correlation is found between SSFR and
age.  In the bottom-left panel, the rates measured in our sample are
compared with the general trend of star-forming galaxies from
\citet{Brinch00} at $z\sim 0.6$ (solid line) and $z\sim 0.9$ (dashed
line).

%%%%%%%%%%%%%%%%%%%%%%%%%%%%%%%%%%%%%%%%%%%
\section{Conclusions}

By comparing optical to NIR photometry of a sample of disc galaxies at
intermediate redshift, with a simple set of phenomenological one-zone
models including chemical enrichment, we are able to explore the
formation of these systems and the evolution of the Tully-Fisher
relation. We find a correlation between the parameters that describe
the star formation history and stellar mass. Dynamical mass
  provides equivalent, but weaker trends. The formation epoch is
especially correlated with stellar mass, showing the usual downsizing
trend, with the most massive discs forming around $z\sim 3$. The
  SF and enrichment timescales have weaker, but detectable
  correlations with stellar mass, towards shorter SF timescales
  ($\tau_1\!\sim\!4$\,Gyr), and fast enrichment times
  ($\tau_2\!\sim\!1$\,Gyr) for the most massive galaxies, reflecting a
  very efficient process for the build-up of metallicity in these
  systems. These are the typical enrichment timescales for the
enrichment of massive early-type galaxies \citep[see, e.g.][]{FS:01},
however, our massive discs have longer SF timescales, resulting in an
extended distribution of stellar ages. On the Tully-Fisher plane, age
does not introduce a segregation at fixed circular speed, suggesting
mild evolution of the Tully-Fisher relation with redshift.

%%%%%%%%%%%%%%%%%%%%%%%%%%%%%%%%%%%%%%%%%%%%%%%%
%%%%%%%%%%%%%%%%  Figure 10   %%%%%%%%%%%%%%%%%%
%%%%%%%%%%%%%%%%%%%%%%%%%%%%%%%%%%%%%%%%%%%%%%%%
%%%%%%%%%%%%%%%%%%%%%%%%%%%%%%%%%%%%%%%%%%%%%%%%

We also present the model fits with respect to dynamical mass
(measured within three scale lengths) and find a similar trend as with
respect to stellar mass. However, the ratio of the two -- an estimator
of dark matter fraction, or feedback efficiency -- seem not to
correlate with the model parameters, reinforcing the idea
that mass is the strongest, first order driver of the star formation
histories of galaxies. Nevertheless, on the Tully-Fisher diagram
(Fig.~\ref{fig:TFz}), at fixed $M_s/M_{\rm dyn}$, older galaxies have
faster circular speeds, perhaps reflecting the fact that older discs
would have formed in higher density halos, hence the higher $V_{\rm
  MAX}$. The difference is rather small, and a more detailed analysis
-- based on spectroscopic data with high SNR and accurate flux
calibration -- is needed to confirm this trend.

Our model includes dust as a free parameter, treated as a simple dust
screen. Our findings agree independently with the formalism of
\citet{tf85} for the most massive discs (stellar mass $\simgt 10^{10}M_\odot$),
with lower mass discs featuring no significant trend between colour
excess and inclination.

%%%%%%%%%%%%%%%%%%%%%%%%%%%%%%%%%%%%%%%%%%%%%%%%
\section*{Acknowledgments}
AB thanks the Austrian Science foundation FWF for funding (project
P23946-N16). The referee is warmly thanked for his/her useful comments
and suggestions. The authors acknowledge the use of the UCL Legion
High Performance Computing Facility (Legion@UCL), and associated
support services, in the completion of this work.

%%%%%%%%%%%%%%%%%%%%%%%%%%%%%%%%%%%%%%%%%%%%%%%%
%%%%%%%%%%%%%%%%  Figure 11  %%%%%%%%%%%%%%%%%%%
%%%%%%%%%%%%%%%%%%%%%%%%%%%%%%%%%%%%%%%%%%%%%%%%
%%%%%%%%%%%%%%%%%%%%%%%%%%%%%%%%%%%%%%%%%%%%%%%%

%%%%%%%%%%%%%%%%%%%%%%%%%%%%%%%%%%%%%%%%%%%%%%%%
%%%%%%%%%%%%%%%   REFERENCES   %%%%%%%%%%%%%%%%%
%%%%%%%%%%%%%%%%%%%%%%%%%%%%%%%%%%%%%%%%%%%%%%%%

\label{lastpage}
\end{document}